\documentclass[prd,preprintnumbers,twocolumn,superscriptaddress,nofootinbib]{revtex4-1}
\pdfoutput=1

\usepackage[utf8]{inputenc}
\usepackage[T1]{fontenc}
\usepackage{amsmath}
\usepackage{float}
\usepackage{graphicx}
\usepackage{comment}
\usepackage[dvipsnames]{xcolor}
\usepackage{xcolor}
\usepackage{color}
\usepackage{epsfig}
\usepackage{dcolumn}
\usepackage{hyperref}
\usepackage{bm}
\usepackage{amssymb}
\usepackage{slashed}
\usepackage{epstopdf}
\usepackage[normalem]{ulem}	
\usepackage{ae}
\pdfoutput=1

\makeatletter
\renewcommand{\maketag@@@}[1]{\hbox{\m@th\normalsize\normalfont#1}}%
\makeatother

\definecolor{red}{rgb}{1,0,0}
\definecolor{darkpink}{rgb}{0.8,0,0.4}

\DeclareRobustCommand{\Eq}[1]{Eq.~(\ref{#1})}
\DeclareRobustCommand{\Fig}[1]{Fig.~\ref{#1}}

\hypersetup{
colorlinks=true,
urlcolor=corlinks,
linkcolor=corlinks,
menucolor=cormenu,
citecolor=corlinks,
pdfborder= 0 0 0
}

\usepackage{tikz,pgfplots}
\usetikzlibrary{arrows,shapes,trees,calc,positioning,patterns}

\definecolor{corlinks}{RGB}{0,0,128}
\definecolor{cormenu}{RGB}{0,0,128}
\definecolor{corurl}{RGB}{0,0,128}

\newcommand{\ew}{\textnormal{\tiny EW}}

\newcommand{\keV}{\,\text{keV}}
\newcommand{\MeV}{\,\text{MeV}}
\newcommand{\GeV}{\,\text{GeV}}

\allowdisplaybreaks 

\begin{document}

\title{Probing photophobic (rel)axion dark matter }

\author{Nayara Fonseca}
\email{nfonseca@ictp.it}
\affiliation{Abdus Salam International Centre for Theoretical Physics,
Strada Costiera 11, 34151, Trieste, Italy}

\author{Enrico Morgante}
\email{emorgant@uni-mainz.de}
\affiliation{PRISMA$^+$  Cluster  of  Excellence  and  Mainz  Institute  for  Theoretical  Physics, Johannes  Gutenberg-Universit\"at  Mainz,  D-55099  Mainz,  Germany}

\begin{abstract}
We investigate the interplay between early universe cosmology and dark matter direct detection, considering axion models with naturally  suppressed couplings to photons.  In the context of the cosmological relaxation  of the electroweak scale, we focus on a scenario of  \emph{Relaxion Dark Matter},  in which the  relaxion field constitutes all the observed dark matter relic density and its allowed mass range is fixed to a few $\keV$ by construction. In particular, we show that a relaxion particle  with  mass   $m_\phi= 3.0 \keV$ which couples to electrons with $g_{\phi, e}= 6.8 \times 10^{-14}$  is consistent with the XENON1T excess, while accounting for the observed dark matter and satisfying astro/cosmo probes. 
This scenario uses the electroweak scale as the link connecting the relaxion production  at early times with the dark matter absorption rate in direct detection.
\end{abstract}

\preprint{MITP/20-042}
\maketitle


\section{Introduction}

In the last decades there has been a huge effort to understand the nature of dark matter (DM). The community has considered different strategies and complementary approaches, such as direct and indirect detection experiments, astro/cosmo probes, and collider searches, but the DM non-gravitational properties are still to be determined. Recently, the XENON1T dark matter experiment has reported an excess of electron recoil events within a noticeable peak  in the $2 \keV$ and  $3 \keV$ energy bins, which contrasts with the expected background model \cite{Aprile:2020tmw}.

The most likely explanation for the excess is a conventional one, such as a statistical fluctuation, a neglected component from Tritium decays, or even another unaccounted background source.  Nevertheless, given the  relevance of the DM question for our understanding of the universe, any viable possibility   deserves an examination.
A solar axion could in principle account for the surplus of events, but the corresponding couplings to electrons and to photons would imply strong tensions with stellar cooling bounds \cite{Aprile:2020tmw} (see also \cite{Viaux:2013lha, Bertolami:2014wua, Giannotti:2017hny}).

As discussed in~\cite{Bloch:2020uzh}, a (pseudoscalar) axion  DM particle with suppressed coupling to photons (``photophobic axion''~\cite{Craig:2018kne}) can provide a good fit to the signal, while escaping bounds from stellar cooling and X-rays emission.
The photophobic relaxion field of~\cite{Hook:2016mqo, Fonseca:2018xzp, Fonseca:2018kqf} is a well motivated example of such a photophobic axion.
The relaxion mechanism is an alternative way of addressing the SM hierarchy problem \cite{Graham:2015cka}, where the value of the weak scale is controlled by the evolution of a classical field in the early universe, the relaxion.
It was shown in Ref.~\cite{Fonseca:2018kqf} that the photphobic relaxion is a viable DM candidate in the $\keV$ range.  In this scenario, relaxation takes place during inflation and the field evolution is stopped due to the backreaction of SM gauge bosons production \cite{Hook:2016mqo}, instead of through the backreaction of a Higgs-dependent barrier \cite{Graham:2015cka}, without the need of new physics at the TeV scale. The relaxion abundance is produced after reheating via the freeze-in mechanism through scatterings of SM particles in the thermal bath, while the misalignment energy is suppressed as the field sets in its potential minimum already during inflation.%
\footnote{See \cite{Banerjee:2018xmn} for a scenario where  relaxion dark matter is produced via coherent oscillations after reheating in a model with Higgs-dependent barriers.}
Constraints from  the cosmic X-ray background on DM decaying into photons can be circumvented  due to the photophobic interactions.
By doing a simple fit to the XENON1T data, we show that, in order to explain the excess, the relaxion must have mass $m_\phi \approx 3.0 \keV$ and a coupling to electrons $g_{\phi,e} \approx 6.8 \times 10^{-14}$, perfectly in line with the predictions of~\cite{Fonseca:2018kqf}.

The structure of this work is the following. In Sec.~\ref{sec:photophobic relaxion} we discuss the  photophobic relaxion, detailing its irreducible couplings to fermions and to photons. We review the results of Ref.~\cite{Fonseca:2018kqf} in Sec.~\ref{sec:review}, where we highlight the   parameter space which is consistent with the excess of electron recoil events reported by the XENON1T experiment. In Sec.~\ref{sec:xenon}, we confront the relaxion dark matter model with  the XENON1T results. We finally summarize our results in Sec.~\ref{sec:conclusions}.

\section{The photophobic (rel)axion}\label{sec:photophobic relaxion}

Let us consider the couplings of an axion $\phi$ with the Standard Model, before electroweak symmetry breaking. Neglecting possible Higgs-portal terms, as well as higher order operators, the Lagrangian reads
\begin{equation} \label{eq:axion general couplings}
\mathcal{L} \supset  g_V^2 \frac{\phi}{4 f_V} V \widetilde{V}  + \frac{\partial_\mu \phi}{f_\psi} (\bar{\psi} \gamma^\mu \gamma_5 \psi),
\end{equation}
where $V$ denotes the SM gauge bosons $V= \{G^a, W^a, B\}$ with the corresponding gauge coupling $g_V$, and  $\psi$ refers to the SM fermions. 
The photophobic axion~\cite{Craig:2018kne} is the case where the coupling to photons is zero at tree level. This is obtained if the UV model satisfies $f_W = -f_B \equiv \mathcal{F}$. As explicitly shown in \cite{Craig:2018kne}, such boundary conditions  may naturally descend from a left-right symmetric model.  Additionally, we assume that in the UV the relaxion does not couple to gluons and fermions, $1/f_G = 1/f_\psi = 0$.
Axion derivative couplings to fermions as in the second term of \Eq{eq:axion general couplings} respect  $\phi$ shift symmetry, allowing them to be generated at lower energies. The renormalization group evolution from  the UV scale $\Lambda$ down to the electroweak scale generates such coupling at 1-loop order as \cite{Bauer:2017ris} (see also \cite{Craig:2018kne}): 
\begin{equation}\label{eq:fF}
 \frac{1}{f_\psi} = -\frac{3\, \alpha_{\textrm{em}}^2 }{4 \mathcal{F}}\left[   \frac{3}{4 \sin^4\theta_W}
- \frac{1}{\cos^4\theta_W}(Y_{\psi_L}^2 +Y_{\psi_R}^2)\right]\log\frac{\Lambda^2}{m_W^2},
\end{equation}
where  $Y_{\psi_{L,R}}$  are the  left and right handed hypercharges of the fermion $\psi$ and $\theta_W$ is the SM weak angle. Furthermore, an irreducible coupling to photons is generated if the axion shift symmetry is explicitly broken due to a mass term. The induced coupling to photons (at one and two loops) is \cite{Bauer:2017ris, Craig:2018kne}:
\begin{equation} \label{eq:fgamma}
\frac{1}{f_\gamma} =  \frac{2\, \alpha_{\rm em}}{ \pi \sin^2\theta_W \mathcal{F}} B_2\left(x_W\right)+  \frac{\alpha_{\rm{em}}}{2 \pi} \sum_\psi \frac{N_c^\psi Q_\psi^2}{2 \pi^2 f_\psi}   B_1\left( x_\psi\right),
\end{equation}
where  $N_c^\psi$ and  $Q_\psi$ are respectively 
the number of colors and the  electric charge of the fermion $\psi$, and $x_i \equiv 4 m_i^2/m_\phi^2$. The functions $B_{1,2}$ are given by:
\begin{align}
B_1(x) = 1 - x [f(x)]^2, \qquad    B_2(x) = 1 - (x - 1) [f(x)]^2 \\
\text{with} \qquad f(x)=\begin{cases}
    \arcsin \frac{1}{\sqrt{x}},   &x \geq 1\\
    \frac{\pi}{2} + \frac{i}{2} \log \frac{1 + \sqrt{1-x}}{1 - \sqrt{1-x}} ,  &x < 1.
  \end{cases}
\end{align}
In the limit where the axion is light  $m^2_\phi\rightarrow 0$, these functions scale like $B_1(x_\psi)\rightarrow -m_\phi^2/(12 m_\psi^2)$ and $B_2(x_W)\rightarrow m_\phi^2/(6 m_W^2)$. As a consequence, the induced coupling to photons in \Eq{eq:fgamma} is  suppressed if the axion is lighter  than the electron.

The relaxion model of Ref.~\cite{Hook:2016mqo}, in which the relaxion evolution is stopped by tachyonic gauge boson production, is built upon the photophobic ALP described above, to which the characteristic relaxion coupling to the Higgs field is added.
The Lagrangian  reads
\begin{align}  \nonumber
\mathcal{L} \supset&  \frac{1}{2} \left(\Lambda^2  -g'\Lambda \phi  \right)h^2 + g \Lambda^3 \phi - \frac{\lambda}{4} h^4  - \Lambda_b^4\cos\left(\frac{\phi}{f'}\right)  \\
& - \frac{\phi}{4\mathcal{F}} \left(g_2^2 W^a_{\mu\nu}\widetilde{W}^{a\,\mu\nu}\! - \!g_1^2 B_{\mu\nu}\widetilde{B}^{\mu\nu} \right),
\label{eq:relaxion_Lagrangian pp}
\end{align}
where $\Lambda$ is the cutoff of the theory,  $g$ and $g'$ are   spurions that explicitly break $\phi$ shift symmetry,  $h$ is the Higgs field, and $\lambda$ is the Higgs  quartic coupling.   $B$ and $W$  are the SM gauge bosons with $g_1$ and $ g_2$ as the  corresponding $\textrm{U}(1)$ and $\textrm{SU}(2)$ gauge couplings. The term in the second line is responsible for the tachyonic production of gauge bosons, as we are going to detail below.
In this model, the effective scale $\mathcal{F}$  is not the same as the axion decay constant $f'$. In this respect, there are different  model building strategies such as alignment mechanism or multi-axion mixings (for different  possibilities see e.g. \cite{Kim:2004rp, Dvali:2007hz, Choi:2014rja, Bai:2014coa, Choi:2015fiu, Kaplan:2015fuy, Fonseca:2016eoo}). The scale $\Lambda_b$ multiplying the cosine  is related to the non-perturbative dynamics  of a new  non-abelian gauge group which gives rise to the periodic potential. The amplitude of these barriers is independent of the Higgs VEV, implying that the relaxion-Higgs mixing   is due to the second term in \Eq{eq:relaxion_Lagrangian pp}, and is given by \cite{Fonseca:2018xzp, Fonseca:2018kqf}:
\begin{equation} \label{eq:mixing pp}
\sin{2\theta_{\phi,h}} = \frac{2  g' \Lambda  v_\ew}{\sqrt{4 g'^2 \Lambda^2 v_\ew^2 + (m_h^2 - m_\phi^2)^2}} \,.
\end{equation}
In the next section, we review the conditions this model should satisfy to make  the relaxion constitutes the observed dark matter density \cite{Fonseca:2018kqf}.

Let us now summarize how the relaxion mechanism is implemented in the model described above.
The starting point is \Eq{eq:relaxion_Lagrangian pp}, in which we expand the last term in mass eigenstates:
\begin{align}\label{eq:lagbroken}
 - \frac{\phi}{\mathcal{F}} \epsilon^{\mu\nu\rho\sigma}& \Big(
2 g_2^2 \partial_\mu W^-_\nu \partial_\rho W^+_\sigma +
(g_2^2-g_1^2) \partial_\mu Z_\nu \partial_\rho Z_\sigma \nonumber \\
& - 2g_1 g_2 \partial_\mu Z_\nu \partial_\rho A_\sigma
\Big)\,.
\end{align}
In our analysis we only consider the tachyonic instability from the $Z\widetilde{Z}$ term above. This simplification is justified by the following two reasons. First, we expect that the term with $W$ bosons is subdominant due to the $W$ self-interactions, which can lead to an effective mass  suppressing  particle production. Second, as discussed earlier, the coupling $\phi F \widetilde{F}$ with photons, the one responsible for tachyonic production, is suppressed by construction due to the structure of the 5-dimensional operator in \Eq{eq:relaxion_Lagrangian pp}. We then expect  that the term $ZA$ in \Eq{eq:lagbroken} to be suppressed compared to the $ZZ$ term.
 Following the convention in \cite{Fonseca:2018kqf}, we absorb the gauge coupling combination multiplying the $\phi Z\widetilde{Z}$ term in the definition of the scale $f$:
\begin{equation} \label{eq:rescaled f}
\frac{1}{f} = \frac{(g_2^2 -g_1^2)}{\mathcal{F}}.
\end{equation}

In the scenario explored here, relaxation  happens during the inflationary epoch. 
Moreover, we assume that the Higgs mass term is initially negative and of the order of the cutoff $\Lambda$. Thus, electroweak symmetry is broken, and the $Z, W$ bosons have a mass close to the cutoff. The relaxion field rolls down the potential thanks to the linear term in \Eq{eq:relaxion_Lagrangian pp}, scanning then the Higgs mass parameter, which controls the value of the Higgs VEV and the mass of the gauge bosons. As $m_Z$ approaches zero,  tachyonic particle production of gauge bosons starts, dissipating the relaxion kinetic energy and making it stop in one of the barriers of the cosine potential. Considering the Lagrangian in (\ref{eq:relaxion_Lagrangian pp}), one can see this backreaction mechanism directly from the equations of motion for $\phi$ and $Z$:
\begin{align}
\ddot\phi - g\Lambda^3 + g'\Lambda h^2 + \frac{\Lambda_b^4}{f'}\sin\frac{\phi}{f'} + \frac{1}{4f}\langle Z\widetilde{Z} \rangle &= 0, \label{eq:phieom} \\
\ddot Z_\pm +(k^2 + \left(m(h)\right)^2 \mp k\frac{\dot\phi}{f})Z_\pm &= 0, \label{eq:Zeom}
\end{align}
with $m(h)=\sqrt{g_1^2+g_2^2}h/2$.   $Z_\pm $ refers to the two transverse polarizations of the field $Z_\mu$  and $\langle Z\widetilde{Z} \rangle$ is the expectation value of the quantum operator, which can be written as:
\begin{equation}
\langle Z\widetilde{Z} \rangle = \int\frac{d^3k}{(2\pi)^3}\left(|Z_+|^2-|Z_-|^2\right) \,.
\end{equation}
Note that we neglected in the equations above the longitudinal component $Z_L$ as it does not have a tachyonic instability in its equation of motion. From \Eq{eq:Zeom} and assuming positive velocity $\dot{\phi}$, $Z_+$ has a mode $k$ with tachyonic growth as soon as  $ \omega_{k,+}^2 \equiv k^2 + \left(m(h)\right)^2 - k\frac{\dot{\phi}}{f} <0$.  
The first mode to become tachyonic is the one for which $ \omega_{k,+}^2$ is minimum, $k=\dot{\phi}/(2f)$.
Consequently, the field $Z_+$ has an exponential growth for 
\begin{equation}\label{eq:tachyonic growth starts}
\dot{\phi} > 2 f\, m_Z(h).
\end{equation}
This dissipation makes the relaxion slow down until its velocity cannot overcome the cosine barriers, and then $\phi$ is trapped in one the wiggles (see e.g. \cite{Fonseca:2018xzp} for a numerical  example).
The electroweak scale is fixed by the point in time at which \Eq{eq:tachyonic growth starts} is satisfied. Imposing that this happens at $m_Z(h) = m_Z \approx 90\GeV$, the scale $f$ can be rewritten in terms of the other parameters in the model as
\begin{equation}\label{eq:scale f}
f = \frac{\dot\phi}{2 m_Z} = \frac{g \Lambda^3}{6 H_I m_Z} \,,
\end{equation}
where we assumed a slow-roll velocity $\dot\phi= g\Lambda^3/(3 H_I)$ with $H_I$  as the Hubble rate during inflation. 
We note that the initial velocity should be larger than the cosine barriers in \Eq{eq:relaxion_Lagrangian pp},  $\dot{\phi} \gtrsim \Lambda_b^2$, so that the relaxion field is able to overcome the wiggles during the scanning process. On the other hand, these barriers should be high enough to stop the field once particle production has turned on. The potential should then have local minima which requires $\Lambda_b^4 \gtrsim g\, \Lambda^3 f'$.

After the tachyonic production of $Z$ bosons starts, these rapidly thermalize, making the discussion more complicate. On the one hand, the large thermal mass of the Higgs temporarily restores the EW symmetry, making the vector bosons light and the tachyonic growth more efficient. On the other hand, the inclusion of a Debye mass for the $Z$ suppresses its further production. These  effects must be taken into account in order to correctly determine the parameter space for the model, for which we refer to Refs.~\cite{Fonseca:2018xzp, Fonseca:2018kqf}.

For this backreaction mechanism to work, it is crucial to guarantee that the $\phi F \widetilde{F}$ coupling to photons is subdominant. If this is not satisfied, tachyonic photon production would be  active during the whole evolution, effectively dissipating the relaxion kinetic energy when the Higgs VEV is still large.  This feature makes the connection to the photophobic  model  described before, which can accomplish such requirement by having maximally suppressed coupling to photons.

In order to get the correct value of the electroweak scale from this relaxion model, a number of conditions should be fulfilled. We refer the reader to Ref.~\cite{Fonseca:2018kqf} for a detailed discussion of such requirements. 
Following Ref.~\cite{Fonseca:2018kqf}, we fix three different ratios of the  parameters $g$ and $g'$  in \Eq{eq:relaxion_Lagrangian pp}, namely $g/g'= 1, 10^3, 10^6$. 
Note that the couplings $g$ and $g'$ have to satisfy the condition  $g>g'/(4\pi)^2$. If this is not the case, the slope term $g\Lambda^3\phi$ in the  potential would be subdominant compared to a linear term  $g'\Lambda^3\phi / (4\pi)^2$ generated via a Higgs loop. 
Although these $g/g'$ ratios  are technically natural choices,  we point out that if these terms are generated in a similar way in  the UV model, one could expect  $g \sim g'$. On the other hand, as shown  in \cite{Fonseca:2018kqf}, relaxing this assumption opens the parameter space, so we  include the three benchmarks in this study.
As we will show in the following, a large ratio $g/g'\sim 10^6$ is necessary to fit the XENON1T excess.
In the left column of Tab.~\ref{tab:range}, we report the allowed range of the parameters of the relaxion model for the benchmark $g/g'= 10^6$.

\begin{table}
\renewcommand{\arraystretch}{2}
\begin{tabular}{|c|c|c|}
\hline
$g/g'$ & Relaxion DM & XENON1T \\ \hline

\scriptsize  $ \displaystyle \frac{\Lambda}{\mathrm{GeV}}$ & \scriptsize $7\cdot 10^3  - 3\cdot 10^6$ & \scriptsize $2.5\cdot 10^5 - 1.4\cdot 10^6$ \\

\scriptsize $  \displaystyle g'$ & \scriptsize $ 3\cdot 10^{-22} - 3\cdot 10^{-16}$ & \scriptsize $ 1\cdot 10^{-19} - 2\cdot 10^{-18}$ \\

\scriptsize $\displaystyle \frac{m_\phi}{\mathrm{keV}}$ & \scriptsize $2 - 17$ & \scriptsize $2 - 4.2$ \\

\scriptsize $ \displaystyle \frac{f'}{\mathrm{GeV}}$ & \scriptsize $3\cdot 10^{10}-3\cdot 10^{16}$ & \scriptsize $1\cdot 10^{15}-3\cdot 10^{16}$ \\

\scriptsize $ \displaystyle \frac{\Lambda_b}{\mathrm{GeV}}$ & \scriptsize $3\cdot 10^2 - 4 \cdot 10^5 $ & \scriptsize $6\cdot 10^4 - 3 \cdot 10^5 $ \\

\scriptsize $\displaystyle \frac{f}{\mathrm{GeV}}$ & \scriptsize $3\cdot 10^5 - 2\cdot 10^9$ & \scriptsize $3\cdot 10^8 - 8\cdot 10^8$ \\

\scriptsize $\displaystyle \frac{H_I}{\mathrm{GeV}}$ & \scriptsize $2\cdot 10^{-11} - 4\cdot 10^{-6}$ & \scriptsize $3\cdot 10^{-8} - 1\cdot 10^{-6}$ \\

\scriptsize $\displaystyle \frac{T_0}{\mathrm{GeV}}$ & \scriptsize $10^{-3} - 3\cdot 10^1$ & \scriptsize $1 - 3\cdot 10^1$ \\

\scriptsize $  \displaystyle N_e$ & \scriptsize $3\cdot 10^{5} - 5\times 10^{12}$ & \scriptsize $3\cdot 10^{5} - 2\times 10^{8}$\\

\scriptsize $  \displaystyle \theta$ & \scriptsize $1\cdot 10^{-18} - 5 \cdot 10^{-14}$ & \scriptsize $1\cdot 10^{-15} - 8 \cdot 10^{-15}$ \\

\hline

\end{tabular}
\caption{\label{tab:range} Allowed parameter space for $g/g' = 10^6$. The last two lines show the minimal number of e-folds of inflation that allows for relaxation to complete, and the value of the relaxion-Higgs mixing angle. Left, parameter space of the Relaxion DM model~\cite{Fonseca:2018kqf}.   Right, parameter space consistent with the $2\sigma$ region off the best-fit point of the XENON1T excess (see Sec.\,\ref{sec:xenon}).}
\end{table}

An important concern about this model comes from the fragmentation of the relaxion field due to its periodic potential~\cite{Fonseca:2019ypl, Fonseca:2019lmc}. When the relaxion field rolls over the periodic barriers, fluctuations are sourced by a parametric resonance. The gradient energy grows at the expenses of the relaxion kinetic energy, and the field possibly slows down until it stops due to the finite size of the barriers well before the critical point $m_h^2\sim 0$ is reached, spoiling the mechanism. As it was shown in Ref.~\cite{Fonseca:2019lmc}, this is not the case in the present scenario, since the amplitude of the barriers $\Lambda_b$ is small enough to suppress the growth of fluctuations.

\section{Photophobic (rel)axion as dark matter}
\label{sec:review}

In this section we review the photophobic relaxion dark matter scenario of  Ref.\,\cite{Fonseca:2018kqf}.    We focus on the impact of the axion irreducible couplings to fermions  and to photons  on the dark matter production  and decays in the early universe.

Since relaxation happens during an inflationary period, the vacuum misalignment contribution to the relaxion abundance is negligible as it is diluted away thanks to the expansion.%
\footnote{
It is interesting to note that, if a photophobic axion has to explain the XENON1T anomaly, its coupling to the SM unavoidably generates a warm axion population. Thus, assuming that most of the DM is generated through the misalignment mechanism would be possible only at the price of having a small reheating temperature $T_0 \lesssim 100 \GeV$ (assuming $\phi$ couples to all SM fermions)~\cite{Takahashi:2020bpq}.}
A population of relaxion particles is produced through scatterings with SM particles. The process proceeds out of equilibrium and a low reheating temperature is necessary to obtain the correct relic abundance. The dominant production channel is, above the QCD scale, the axion-gluon Compton scattering $g + q \leftrightarrow \phi + q$,  and below the QCD scale the axion-photon Compton scattering $\gamma + \psi \leftrightarrow \phi + \psi$, both mediated by the axion coupling to fermions of Eq.~(\ref{eq:fF}). The rates for these processes are given by
\begin{align}
\Gamma_{C,\gamma} & = \frac{3\zeta(3)}{\pi^2}\alpha_\mathrm{em} \frac{m_\psi^2 T}{f_\psi^2} \,, \label{eq:rate Compt photons}\\
\Gamma_{C,g} & = \frac{36\zeta(3)}{\pi^2}\alpha_s \frac{m_\psi^2 T}{f_\psi^2} \,,\label{eq:rate Compt gluons}
\end{align}
where $\alpha_\mathrm{em}$ is the fine structure constant, $\alpha_s$ is the  QCD strong coupling, and $\zeta(3)\approx 1.2$.
Due to the fermion-mass suppression of this coupling, the fermion contributing the most to the axion production is, at any given temperature, the most massive one which is still relativistic (and thus not Boltzmann-suppressed).
The coupling of the photophobic axion to fermions has two sources. One is the loop-induced coupling of Eq.~(\ref{eq:fF}). If the axion is the relaxion, an additional coupling descends from its mixing with the Higgs, Eq.~(\ref{eq:mixing pp}). In the case of relaxion DM, though, the mixing angle is suppressed in such a way that the axion-fermion coupling is dominated by Eq.~(\ref{eq:fF}). This implies that the relation among the couplings to SM particles is completely fixed by the photophobic nature of the axion, and it is independent of the relaxion properties. In particular, the same scale $f$ controls the production of relaxion particles independently of the dominant production channel, through the dependence of $f_\psi$ on $f$. Consequently, our results are valid in the generic case of a photophobic axion.
The only dependence on the properties of the relaxion mechanism lies in the allowed range of $f$ and consequently $g_{\phi,e}$.
Indeed, the scale $f$ controlling the strength of particle production  is fixed in terms of the other parameters in \Eq{eq:scale f}  in which we define the eletroweak scale. Both the axion production rate at early times and the axioelectric absorption cross section in a dark matter detector are related by fixing the electroweak scale at the correct value. This is a key difference compared to a generic photophobic axion. In the latter case, as long as cosmological and astrophysical bounds are satisfied,  one can freely adjust the scale controlling the coupling $1/f_e$ in (\ref{eq:fF}).%
\footnote{Note that the scale $\mathcal{F}$ in \Eq{eq:fF} is just a  rescaling of $f$, see (\ref{eq:rescaled f}).}

The XENON1T results can be expressed in terms of the the effective dimensionless coupling of the photophobic axion to electrons, defined as
\begin{equation} \label{eq:g_phi,e}
g_{\phi, e} = - \frac{2 m_e}{f_e}.
\end{equation}
The scale $f_e$ is related, through Eq.~(\ref{eq:fF}), to the scale $f$ and to all the other couplings $1/f_\psi$. Thus, by requiring that the measured DM abundance is matched, given the reheating temperature the coupling $g_{\phi,e}$ is uniquely determined. 
In Fig.~\ref{fig:mphi x gphi,e} we show, for the three benchmarks $g/g'=1, 10^3, 10^6$, the value of $g_{\phi, e}$ that is required to match the observed DM abundance, for different values of the reheating temperature $T_0$. The choice of values of $T_0$, and the corresponding ranges of $m_\phi$, is made in order to satisfy the conditions discussed in Ref.~\cite{Fonseca:2018kqf} to guarantee a successful relaxation of the EW scale, while passing the indirect detection X-rays constraints. For $g/g'=1$ and $g/g'=10^3$ the reheating temperature cannot be larger than $\sim 100\, \mathrm{MeV}$. 
Consequently,
the relaxion-electron  couplings need to be large and are in tension with   XENON1T results, which point to a coupling  $g_{\phi, e} \sim \mathrm{few} \times 10^{-14}$. 
 Since the $g/g'= 10^6$ case allows for  higher values of $T_0$ and thus of $f_e$ in \Eq{eq:fF}, it  can reach the  $g_{\phi, e}$ coupling necessary to match the XENON1T signal.

\begin{figure}
\center
\includegraphics[width=0.49\textwidth]{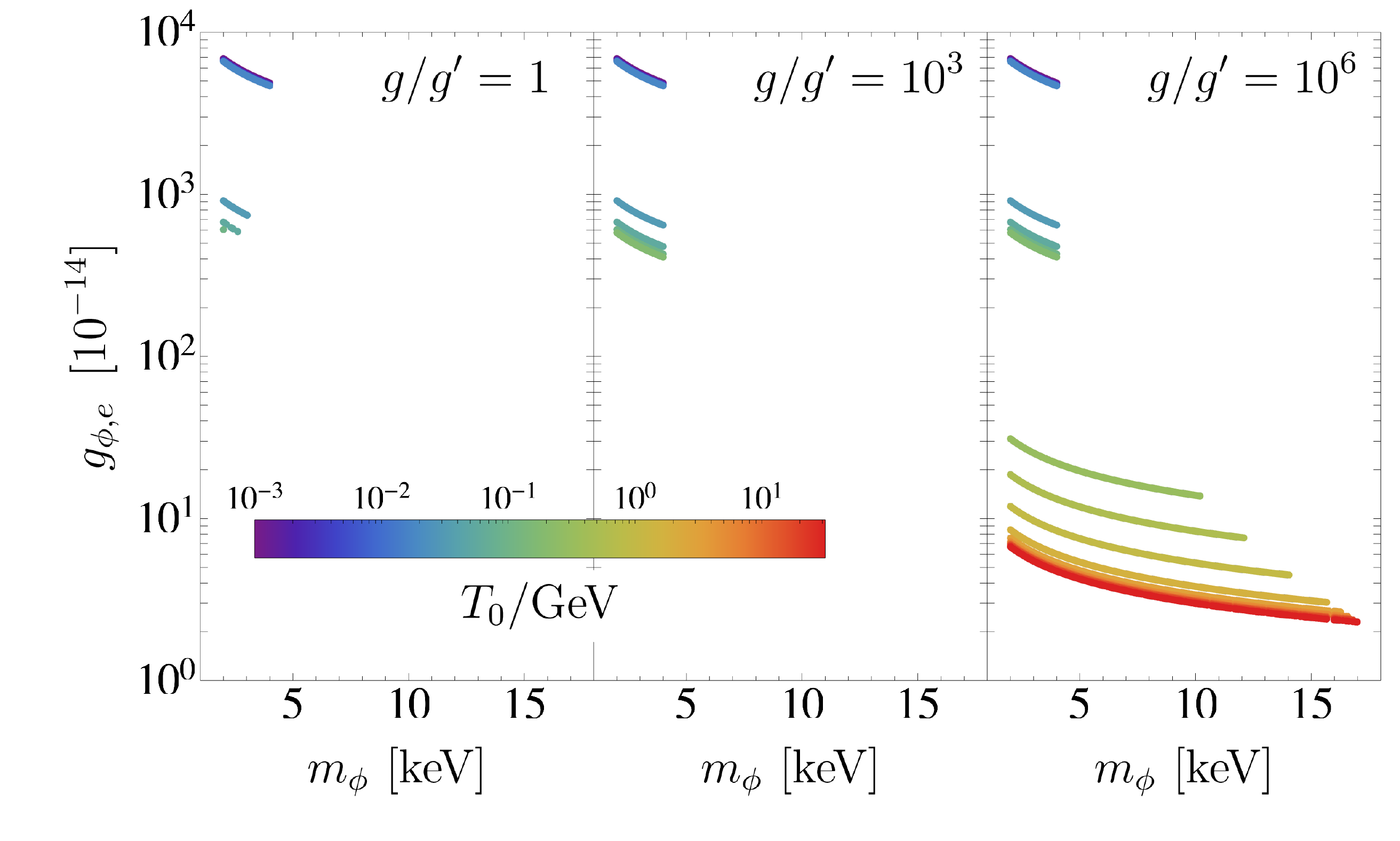}
\caption{Allowed points in the $(m_\phi, g_{\phi,e})$ plane where relaxion can account for the whole observed DM density while fulfilling   relaxation requirements to set the correct value of the electroweak scale and satisfying X-ray constrains. We show the three scenarios discussed in the text, $g/g'= 1, 10^3, 10^6$. $T_0$ is the corresponding reheating temperature. Note that for the highest values of $g_{\phi, e}$ some points overlap for  different scenarios. The case $g/g'=10^6$ can match the  XENON1T excess in eletron recoil events as discussed in Sec.\,\ref{sec:xenon}.}
\label{fig:mphi x gphi,e}
\end{figure}

We assume that the relaxion relic abundance before reheating is negligible. A DM population is then built up from the time the universe starts to redshit as radiation. In order to avoid overabundance, the reheating temperature  for the scenario $g/g'=10^6$ can vary between $\sim 1 \MeV - 30 \GeV$~\cite{Fonseca:2018kqf}. Note that the production rates in Eqs.~(\ref{eq:rate Compt photons}, \ref{eq:rate Compt gluons}) scale with the second power of the axion-SM couplings, such that the highest temperatures in the  allowed range  are reached for the smallest values of $1/f$, and correspondingly for  the smallest values of $g_{\phi,e} $  in \Fig{fig:mphi x gphi,e}, which is the preferred region for the XENON1T signal.

 As shown in Ref.~\cite{Fonseca:2018kqf}, the relaxion mass  for $g/g'=10^6$ can vary from $2$ to $17 \keV$. The lower bound is imposed due to constraints from structure formation.
Measurements from Lyman-$\alpha$   constrain thermal relics  lighter than  a few $\keV$ \cite{Viel:2013apy,Irsic:2017ixq}. We point out that in our scenario the DM is produced out-of-equilibrium such that its velocity distribution can depart from a thermal one, which can relax these bounds (see e.g. \cite{Bernal:2017kxu,Heeck:2017xbu}). In addition, if warm DM only constitutes a fraction of the total DM abundance, bounds from  Lyman-$\alpha$ can be alleviated, see e.g. \cite{Takahashi:2020bpq}.
In particular, if the relaxion composes a fraction $x<1$ of DM, our results would change only for a rescaling of the axion-electron coupling by a factor of $(1/x)^{1/2}$ in order to match the XENON1T signal (see next section), and a rescaling of the reheating temperature by another factor $x^2$ to match the relic abundance. A fraction $x\approx 10\%-20\%$ would evade structure formation limits even for sub-keV masses, without being excluded by the constraints on the relaxion model.
The resulting reheating temperature for $g/g'=1,10^3$ would be at most $\mathcal{O}(1)\MeV$, in tension with Big Bang nucleosynthesis.

In the relevant mass range, the relaxion can only decay into photons and neutrinos.  As in \cite{Fonseca:2018kqf}, we assume that the decay into neutrinos is suppressed by considering the case in which neutrinos are Majorana fermions, see e.g. \cite{Bauer:2017ris}. Despite this, indirect detection constraints from X-rays, accounted for in this work, impose even stronger bounds on the relaxion DM lifetime, and therefore the case of Dirac neutrinos does not add any constraints to the model.
The relaxion decay into photons happens through the mixing with the Higgs in Eq.~(\ref{eq:mixing pp}) and via the loop-induced coupling $1/f_\gamma$ in Eq.~(\ref{eq:fgamma}).  For the case $g/g'=1$, the coupling $g'$ reaches higher values such that there is a region of the parameter space where the decay into photons via mixing can dominate, while for the cases $g/g'=10^3$ and  $g/g'=10^6$ the decay through the coupling $1/f_\gamma$  always dominates.

\section{XENON1T signal and relaxion dark matter}\label{sec:xenon}

We consider the case in which the excess observed by the XENON1T detector is attributed  to  relaxion absorption via axioelectric effect rather than due to particle-electron scattering. The axioelectric absorption cross section is related to the
photoelectric cross section as~\cite{Dimopoulos:1986mi, Dimopoulos:1985tm, Pospelov:2008jk}
\begin{equation}
\sigma_{\rm{\phi e}} (E_\phi) = \sigma_{\rm{pe}}(E_\phi) \frac{ g_{\phi,e}^2}{v_\phi} \frac{3\,E_\phi^2}{16 \pi \alpha_{\rm{em}} m_e^2}\left(1 - \frac{v_\phi^{2/3}}{3}\right),
\end{equation}
where $E_\phi$ is the axion total energy, $g_{\phi,e}$ is the dimensionless axion-electron coupling, $v_\phi$ is the axion velocity, $\alpha_{\rm{em}}$ is the fine structure constant, and $m_e$ is the electron mass. We took the photoelectric cross section $\sigma_{\rm{pe}}$ from the database in Ref.~\cite{XCOM}. Assuming that the axions are non-relativistic ($v_\phi\ll 1$) and constitute the local DM density, the predicted signal  is a mono-energetic peak at the axion rest mass. The expected spectrum is then a smeared peak due to the limited detector resolution.  The differential event rate of axion dark matter absorption per unit of energy  in  the XENON1T experiment  is given by
\begin{equation} \label{eq:dR/dE}
\frac{dR}{dE} = \Phi^{\rm{DM}} \sigma_{\rm{\phi e}}(E) \delta(E -m_\phi),
\end{equation}
where  $\Phi^{\rm{DM}} = \rho_{\rm{DM}} v_\phi/m_\phi $ is the DM flux with $\rho_{\rm{DM}}= 0.4 \GeV \cdot \rm{cm}^{-3}$ being the local DM density.
From \Eq{eq:dR/dE} one can  obtain the total number of events for the XENON1T energy range by convoluting the mono-energetic expected signal with the detector resolution, considering the detector efficiency and the total exposure of 0.65 tonne-year  \cite{Aprile:2020tmw}. For the detector energy resolution, the theoretical prediction in \Eq{eq:dR/dE} was smeared assuming a Gaussian distribution with the energy dependence of Ref.\,\cite{Aprile:2020tmw}.  As discussed in Sec.\,\ref{sec:review}, the relaxion model with  ratio $g/g'=10^6$  can match  the  XENON1T excess in electronic recoil events.  In this case,  the relaxion interactions  with electrons  and photons through the mixing with the Higgs are  subdominant compared to the loop-induced couplings originated from the dimension-5 operator in \Eq{eq:relaxion_Lagrangian pp}. We  consider this benchmark for the best-fit analysis in the following.

\begin{figure}
   \center
    \includegraphics[width=.45\textwidth]{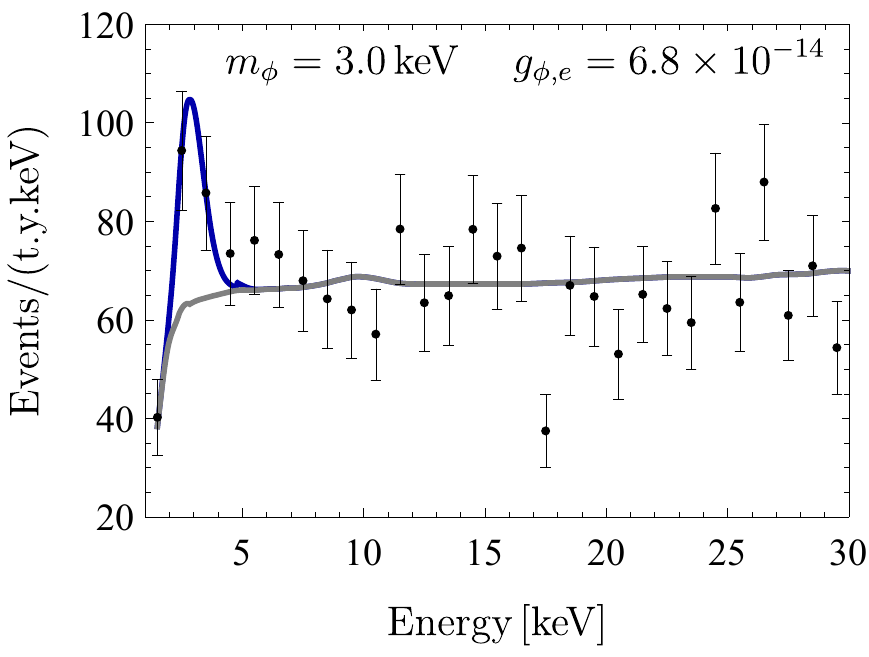}          
    \includegraphics[width=.45\textwidth]{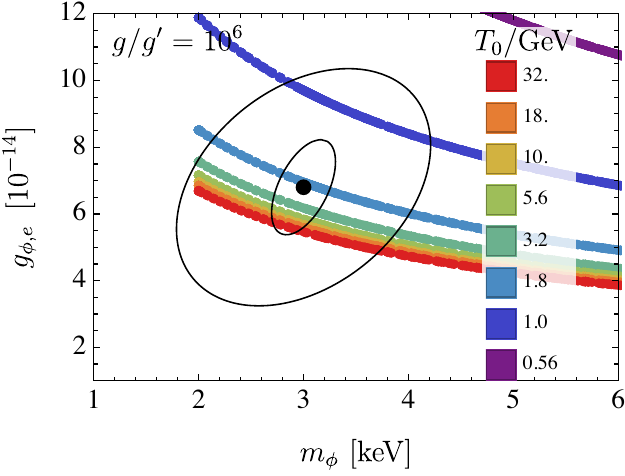}
        \caption{ \label{fig:best-fit} Fits of the XENON1T excess in electronic recoil events  reported in Ref.~\cite{Aprile:2020tmw}. \textbf{Top:} Best-fit + background is shown in blue  with values in Eq.~(\ref{eq:best fit}). The data points and background model (gray) were obtained from Ref.~\cite{Aprile:2020tmw}.  \textbf{Bottom:} Error ellipses in the $(m_\phi \,[\keV], g_{\phi,e}\,[10^{-14}])$ plane correspond to  $1 \sigma$ and $2\sigma$ regions off the best-fit point in black given in Eq.~(\ref{eq:best fit}). The coloured dots are the $\{m_\phi, g_{\phi,e}\}$ points consistent with the relaxion  model with stopping mechanism via gauge boson production, matching the DM observed abundance and satisfying X-ray constraints (corresponding  to  the case  $g/g'=10^6$  in the $(m_\phi, g_{\phi,e})$ plane in Fig.~\ref{fig:mphi x gphi,e} of Sec.~\ref{sec:review}). The $T_0$ values indicate the corresponding reheating temperatures.}
   
\end{figure}

We have digitized the signal and background model from figures in Ref.~\cite{Aprile:2020tmw} at the level of 1-keV-binned data from 1 keV to 30 keV. The reconstructed keV-binned  spectrum considers both scintillation (S1) and  ionization (S2) signals produced by a particle which interacts with the detector~\cite{Aprile:2020tmw}.%
\footnote{We note that the XENON1T S2-only analysis~\cite{Aprile:2019xxb} has a lower energy threshold, reaching higher sensitivity for scenarios where the signal is explained by a sub-keV particle (see e.g. Ref.~\cite{Budnik:2019olh}). }
Using a $\chi^2$ test statistic, we then compare our result to the background prediction $B_0$.  The best-fit signal hypothesis is
\begin{align} \label{eq:best fit}
& m_\phi = 3.0 \keV \quad \quad g_{\phi,e} = 6.8 \times 10^{-14} \nonumber \\
& \chi_{\rm{S+B}}^2 = 35.7\, (27\, \rm{d.o.f.}).
\end{align}
The corresponding spectrum  is shown in the top panel of  Fig.\,\ref{fig:best-fit}.
For the background model only we obtain  $\chi_{\rm{B}}^2 = 46.3\, (29\, \rm{d.o.f.})$.  In the bottom panel of Fig.~\ref{fig:best-fit}, we show the $1 \sigma$ and  $2 \sigma$ regions around the best-fit point in \Eq{eq:best fit}. The coloured dots correspond to the points which are consistent with the relaxion dark matter model and the different values of $T_0$ indicate the corresponding reheating temperatures. It is clear from this plot that our model can well explain the XENON1T data for $T_0\sim 1-30\GeV$. In the right column of Tab.~\ref{tab:range}, we show the allowed range of the parameters of the relaxion model with $g/g'= 10^6$  which are consistent with the $2\sigma$ region off the best-fit point in \Eq{eq:best fit}.

The scale $f$ corresponding to the best-fit point in Eq.~(\ref{eq:best fit})  can be obtained from Eq.~(\ref{eq:fF}) and Eq.~(\ref{eq:rescaled f}) with $f_e = - 2 \,m_e/g_{\phi,e} = - 1.5 \times 10^{10} \GeV$,  and is given by $f= 5 \times 10^8 \GeV$ for a cutoff scale of $\Lambda= 5 \times 10^5 \GeV$. This is in agreement with astrophysical probes constraining photophobic axions \cite{Craig:2018kne}.  The coupling to electrons  is constrained by red giant star cooling bounds, implying  $f\gtrsim 3 \times 10^7 \GeV$. In addition, the  bound from Supernova 1987A can constrain a photophobic axion mainly due to nucleon bremsstrahlung, which results in the lower bound of $f\gtrsim  10^8 \GeV $. We note however that bounds from astrophysical sources are usually associated with uncertainties of about an order of magnitude, see e.g  Ref.~\cite{Krnjaic:2015mbs,Flacke:2016szy,Craig:2018kne}.

\medskip

Before concluding, we note that in the original relaxion case of  Ref.~\cite{Graham:2015cka} $\phi$   couples to photons and to electrons only via the Higgs field through the same mixing angle. This is different from the photophobic relaxion~\cite{Hook:2016mqo}, which besides the coupling with the Higgs, also has  pseudoscalar interactions. 
In the scenarios where the  interaction with gauge bosons in \Eq{eq:relaxion_Lagrangian pp} is absent, the relaxion effective couplings to the SM  are like those of a CP-even scalar, implying that its phenomenology is similar to that of a  Higgs-portal model \cite{Flacke:2016szy, Frugiuele:2018coc}. In fact, this model offers a different scenario to explain the  XENON1T excess, which is attributed to scalar-like relaxions  produced in the Sun.   Reference~\cite{Budnik:2020nwz} investigates this explanation considering  a relaxion model with Higgs-dependent barrier.    In this case, the best-fit parameters of the XENON1T excess are in tension with  stellar cooling bounds.
However, as discussed in Ref.~\cite{Budnik:2020nwz} (see also Ref.~\cite{Bloch:2020uzh}), one can consider the case where these light particles are chameleon-like, such that their production depends on the environment.  In particular, the large densities of red giant stars may destabilize the relaxion shallow potential and locally increase the relaxion mass, which could make possible  to evade stellar cooling bounds.
A possible risk of this scenario is that the relaxion is destabilized and bubbles are formed in which the relaxion rolls down until it reaches a global minimum of its potential, in which the Higgs has, in this model, a large and positive mass term~\cite{Budnik:2020nwz}. Even though this discussion would certainly be interesting, it falls beyond the scope of the present work.

\section{Conclusions}\label{sec:conclusions}

In this work we have interpreted the  XENON1T excess in electronic recoil events as the absorption of  relaxion dark matter particles.  We  considered the dark matter scenario of Ref.~\cite{Fonseca:2018kqf}  in which the relaxion is a photophopic axion-like particle with characteristic couplings to electrons and to photons.  In this scenario, the  relaxion can explain the whole observed dark matter and its allowed mass window is fixed to a few $\keV$. Due to the loop-induced couplings to electrons and to photons, the phenomenology of this pseudoscalar  differs from the usual relaxion model with Higgs-dependent barriers, which connects to the SM only via the mixing with the Higgs field. 
 
We show that a photophobic relaxion dark matter with mass $m_\phi = 3.0 \keV$ which couples to electrons with $g_{\phi, e}= 6.8 \times 10^{-14}$ is consistent with the XENON1T excess, while satisfying astrophysical probes.  In this framework, the  constraints from the cosmic X-ray background can be satisfied due to the naturally suppressed coupling to photons.

We highlight that our results also hold  for the generic case of a photophobic axion. This happens because in our relaxion DM scenario the mixing with the Higgs is suppressed such that the axion-fermion interaction is dominated by the loop-induced coupling in Eq.~(\ref{eq:fF}). Consequently,  the $\phi$ couplings to the SM particles are independent of the relaxion features and  are fixed by the photophobic nature of the axion.  
The only dependence on the relaxion mechanism  lies in the allowed range of the scale that controls the SM gauge boson production, the scale $f$. Such scale is fixed  in terms of the other relaxion model parameters  once we define the eletroweak scale, see \Eq{eq:scale f}. As as result,  $f$ controls  both the axion production rate at early times and the axioelectric absorption cross section in a dark matter detector.
The value of $f$, a free parameter in the generic photophobic construction, is further constrained in the relaxion scenario by fixing the electroweak scale at the correct value.

To conclude, we stress once more that the scenario considered in this work and previously presented in Ref.~\cite{Fonseca:2018kqf}, in which the axion mass range is constrained to a few $\keV$, fits the surplus of events without additional ingredients. This, together with the connection with the hierarchy problem provided by the relaxion mechanism, makes our scenario a minimal explanation for the XENON1T results.

\acknowledgments{
The work of EM is supported by the Cluster of Excellence ``Precision Physics, Fundamental Interactions, and Structure of Matter'' (PRISMA+ EXC 2118/1) funded by  the  German  Research  Foundation (DFG)  within  the German  Excellence  Strategy  (Project  ID  39083149).
}

\bibliography{PPRelDMBib}

\end{document}